\documentstyle[multicol,aps,prb,epsf]{revtex}
\newcommand{\Vec}[1]{\mbox{\boldmath$#1$}}
\begin{document}

\draft

\title{Superconductivity in repulsive electron systems having three-dimensional disconnected Fermi surfaces}

\author{
Seiichiro Onari$^1$, Ryotaro Arita$^1$, Kazuhiko Kuroki$^2$, and Hideo Aoki$^1$
}

\address{$^1$Department of Physics, University of Tokyo, Hongo,
Tokyo 113-0033, Japan}
\address{$^2$Department of Applied Physics and Chemistry,
University of Electro-Communications, Chofu, Tokyo 182-8585, Japan}

\date{\today}

\maketitle

\begin{abstract}
The idea of raising $T_c$ in the 
spin-fluctuation mediated superconductivity on disconnected Fermi surfaces 
with the gap function changing sign across but not within the
Fermi pockets, proposed by Kuroki 
and Arita for two dimensions (2D), is here extended to three-dimensional 
(3D) systems.  Two typical cases of 
3D disconnected Fermi surfaces (stacked bond-alternating lattice and 
stacked ladder layers) are considered.  
By solving Eliashberg's equation for Green's function 
obtained with the fluctuation exchange approximation (FLEX) 
for the repulsive Hubbard model on these structures, we have shown that 
$T_c$ can indeed reach $O(0.01t)$, which is almost 
an order of magnitude higher than in ordinary 3D cases and similar to 
those for the best case found in 2D.  
The key factor found here for the favorable condition for the 
superconductivity on disconnected Fermi surfaces is that 

the system should be quasi-low dimensional, 
and the peak in the spin susceptibility should be 
appropriately ``blurred''.

\end{abstract}


\begin{multicols}{2}
\narrowtext

\section{Introduction}
The monumental discovery of high-$T_c$ 
superconductivity in strongly correlated electrons in 
cuprates\cite{bed-mul} has kicked off intensive studies on 
superconductivity in repulsively interacting electron 
systems.  
These studies have brought to light the spin-fluctuation mediated 
pairing mechanism, which gives rise to anisotropic Cooper pairing 
instabilities.  
So identifying the optimum condition for pairing in the 
repulsively interacting electron systems is one 
of the most fascinating goals of theoretical studies, since it 
should give a guiding principle in searching for 
higher-$T_c$ materials.  

A fundamental question in this context is a repeatedly asked one: 
``why is $T_c$ in the electron-mechanism 
superconductivity so low?"  Although this may at first sound strange, 
this is true in that the estimated $T_c$ for ordinary lattices is at best $O(0.01t)$\cite{arita} 
which is two orders of magnitude smaller than the starting 
electronic (hopping) energy, $t$.  
There are in fact good reasons why $T_c$ is so low.  One is that the 
spin-fluctuation mediated interaction is much weaker than the 
original electron-electron interaction.  
Another important reason, also quite inherent in the 
electron mechanism, is the following.  
If one wants to have superconductivity from repulsion, 
pairing has to be anisotropic, since the nodes in 
the gap function, which convert 
the repulsion to an attraction in the gap equation, realize 
the superconductivity.  Namely, the pair-scattering processes 
(the matrix elements of the interaction that connect 
two electrons around the Fermi surface to another positions) 
contribute positively to the gap equation when there is a 
node across the pair-scattering momentum transfer.  
Unfortunately, the nodes greatly reduce $T_c$:  While 
the gap function has opposite signs across 
the main pair scattering (e.g., $\Vec{k} = (0, \pi) \leftrightarrow 
(\pi, 0)$ for the antiferromagnetic spin-fluctuation exchange), 
there are other pair scattering processes across the 
regions of the Fermi surface having the same sign, 
and these work against the pairing and lowers $T_c$. 
Since the nodes are imperative in anisotropic-pairing superconductivity, 
this problem is unavoidable as far as the ordinary lattices are concerned.

So a novel avenue to explore is: 
can we improve the situation by going over to multiband systems. 
In this context Kuroki and Arita\cite{KA} have proposed that 
the systems that have {\it disconnected 
Fermi surfaces} are a promising candidate for higher-$T_c$ superconductivity.  
In such systems the gap function can change sign {\it across the
Fermi surfaces} but not within each Fermi surface,
so that all of the main pair-scattering processes
contribute positively to superconductivity.
The estimated 
$T_c$ for two-dimensional (2D) repulsive Hubbard model on such lattices is 
indeed almost an order of magnitude higher, $T_c\sim 0.1t$.
The disconnected Fermi surface has been subsequently explored in 
other models and systems\cite{Takashi1,Takashi2,KKA}.  

As for the dimensionality of the system, on the other hand, 
three of the present authors have systematically 
studied\cite{arita} the Hubbard model on 2D and 3D single-band lattices 
with the fluctuation-exchange approximation (FLEX), and have 
clarified that 2D systems are generally more favorable than 3D systems 
as far as the spin-fluctuation-mediated superconductivity 
in ordinary lattices (square, triangular, fcc, bcc, etc) 
are concerned.  The ``best case" among the ordinary 
lattices is identified to be the d-wave pairing in the square lattice, 
and the above-mentioned $T_c\sim O(0.01t)$ in fact refers to this case.  
Physical reason why 2D is more favorable 
is that the fraction of the phase volume over which 
the fluctuation-mediated interaction exerts significant attractions 
is much smaller in 3D.\cite{arita}
$T_c$ estimated by solving Eliashberg's equation 
with FLEX is indeed only $O(0.001t)$ even for the best 3D case 
(as far as simple lattices are concerned).  
Similar results have also been obtained by Monthoux {\it et al.}\cite{mon-lon} 
independently by means of a phenomenological calculation.
Recent experimental finding\cite{rh} that a series of heavy-fermion 
compounds Ce(Rh,Ir,Co)In$_5$ has higher 
$T_c$ when the structure is more two-dimensional (with larger $c/a$) 
is consistent with this prediction 
(although orbital degeneracies in the heavy-fermion 
compound may exert some effects).

\begin{figure}
\begin{center}
\leavevmode\epsfysize=40mm 
\epsfbox{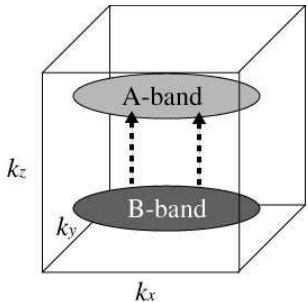}
\caption{Schematic inter-band nesting (arrows) in 
disconnected Fermi surfaces in 3D.}
\label{pri}
\end{center}
\end{figure}

Now, if one puts these two observations (i.e., 
one on the disconnected Fermi surface 
and another on the dimensionality), a natural question arises:  
can we conceive {\it 3D lattices having disconnected Fermi surfaces} 
that have high $T_c$'s.  In other words, 
can the disconnected Fermi surface overcome the 
disadvantage of 3D to possibly realize $T_c\sim O(0.1t)$), 
which should be a fundamental question for the spin-fluctuation 
mediated pairing. 
More specifically, what we have in mind 
is the {\it inter-band nesting} in the 3D disconnected Fermi 
surface as schematically depicted in Fig.\ref{pri}.  
There, a nesting vector exists across the two bands, and this 
is envisaged to give the attractive pair-scattering interaction 
with a nodal plane running in between the two bands.  
Note that a pair is formed within each band, although the nesting is
inter-band. In such a case the gap function should be nodeless within each band, 
while the gap has opposite signs between the two bands.  
In this sense the mechanism is an extension
of the time-honored Suhl-Kondo mechanism, although there are
important differences since here we consider bands
originating from multiple orbitals of the same type
connected by single electron hopping
in order to have appropriate
nesting between the disconnected Fermi surfaces,
while Suhl-Kondo considers basically the bands
originating from different (e.g., s and d) orbitals
that interact only via multi electron processes.

For the survey in lattice structures, there are two possible approaches: 
(i) to find 3D lattice structures 
whose Fermi surfaces are disconnected, and consider the 3D-version of 
the mechanism originally proposed for 2D lattices,\cite{KA}
(ii) stack the 2D lattices already 
proposed\cite{KA,Takashi1,Takashi2,KKA} 
for which $T_c \sim O(0.1t)$ has been estimated, and 
introduce the interlayer hopping, $t_z$, to see 
how the $T_c$ behaves as $t_z$ is increased.  
As for the former approach, we have recently studied superconductivity 
in 3D stacked honeycomb lattice,\cite{onari} where the 
Fermi surface consists of two networks of tubes.  
However, the networks of tubes have turned out to have such a strong 
structure that the intra-band 
as well as inter-band pair-scattering processes contribute to the 
gap equation, and the resulting gap function 
has nodes (with $d$-wave symmetry) within each band.   Accordingly, 
the estimated $T_c \sim O(0.001t)$ is low, and   
the stacked honeycomb lattice does not realize the 
3D-version of inter-band nesting conceived here.

This has motivated us in the present study to look for 
models that have more compact and structure-less Fermi surfaces. 
We have found that a stacked bond-alternating 
lattice (Fig. \ref{lattice}) has a compact and disconnected 
Fermi surface (i.e., a pair of ball-like Fermi pockets).  
We then show for the multiband Hubbard model on this lattice 
that the inter-band pair-scattering alone is dominant.  
The estimated $T_c$ is $O(0.01t)$, 
which is the same order of that for the square lattice, and remarkably 
high for a 3D system.

As for the second approach of stacking the 
2D lattices, we have studied the $t_z$-dependence of 
the pairing instability for stacked ladder lattice (Fig. \ref{lattice}).  
We have found that the $T_c \sim O(0.1t)$, found for the 
single layer, is unexpectedly 
robust against, or even increases with, $t_z$ 
when $t_z$ is not too strong.
We have then identified the underlying physics 
for the favorable condition for the 
superconductivity on disconnected Fermi surfaces --- 
the system should be quasi-low dimensional, 
and the peak in the spin susceptibility should be 
appropriately ``blurred''.

So we shall conclude that the idea of raising $T_c$ in 
superconductivity from repulsive electron-electron interaction 
on disconnected Fermi surfaces 
can work in 3D lattices if we consider appropriate 
lattice structures. 

\section{formulation}
\begin{figure}
\begin{center}
\leavevmode\epsfysize=40mm 
\epsfbox{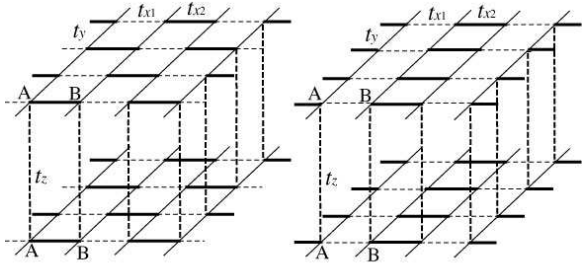}
\caption{A stacked bond-alternating lattice (left panel) and 
a stacked connected-ladder lattice (right) considered in the present 
work.  Solid lines (dashed lines) indicate stronger (weaker) 
hopping, and A, B label sublattices.}
\label{lattice}
\end{center}
\end{figure}

Here we take the repulsive Hubbard model, a simplest possible 
model for repulsively interacting 
electron systems.  So the interaction is the ordinary on-site repulsion, $U$, 
but the lattice has two atoms per unit cell here 
accompanied by two bands.  The Hamiltonian is
\begin{eqnarray}
{\cal H}&=&\sum_{i,j}^{\rm nn}\sum_{\sigma=\uparrow,\downarrow}\sum_{\alpha,\beta={\rm A},{\rm B}}t_{ij}c_{i\sigma}^{\alpha\dagger}c_{j\sigma}^{\beta}\nonumber\\
& &+U\sum_{i}\sum_{\alpha={\rm A},{\rm B}}n_{i\uparrow}^{\alpha}n_{i\downarrow}^{\alpha},
\end{eqnarray}
where $\alpha, \beta$ label A,B sublattices 
with standard notations otherwise.  
For the transfer energies $t$, $t_{x_1}$, $t_{x_2}$ and $t_y$ are 
intra-layer hoppings, while $t_z$ the inter-layer (see Fig. \ref{lattice}).

We employ the FLEX method developed by Bickers
{\it et al}.\cite{bickers1,bickers2,dahm,bennemann}
In the two-band version 
of FLEX,\cite{koikegami,kontani} Green's function $G$, self-energy $\Sigma$, 
susceptibility $\chi$, and the gap function
$\phi$ all become $2 \times 2$ matrices, such as $G_{\alpha\beta}(k)$,
where $\alpha,\beta=$ A,B sublattice, 
$k\equiv(\Vec{k},i\omega_n)$ with $\omega_n=(2n-1)\pi T$ being the
Matsubara frequency for fermions. 

The self-energy is given in the FLEX as
\begin{eqnarray}
\Sigma_{\alpha\beta}(k)
&=&\frac{T}{N}\sum_{k'}\left[V_{\alpha\beta}^{\rm ph}(k-k')G_{\alpha\beta}(k')\right.\nonumber\\
& & \left.-V_{\alpha\beta}^{\rm pp}(k-k')G_{\beta\alpha}(-k')\right],
\end{eqnarray}
where the particle-hole ($V^{\rm ph}$) and particle-particle 
($V^{\rm pp}$) fluctuation-exchange interactions are given as
\begin{eqnarray}
V_{\alpha\beta}^{\rm ph}(q) 
&=&\frac{3}{2}U^2\!\!\left[\frac{\chi^{{\rm ph}}(q)}{1-U\chi^{{\rm ph}}(q)}\right]_{\alpha\beta}\!\!\!+\frac{1}{2}U^2\!\!\left[\frac{\chi^{{\rm ph}}(q)}{1+U\chi^{{\rm ph}}(q)}\right]_{\alpha\beta}\!\!\!\!\nonumber\\
& &-U^2\chi_{\alpha\beta}^{\rm ph}(q),
\end{eqnarray}
\begin{equation}
V_{\alpha\beta}^{\rm pp}(q)=U^2\left[\frac{\chi^{{\rm pp}}(q)}{1+U\chi^{{\rm pp}}(q)}-\chi^{\rm pp}(q)\right]_{\alpha\beta}\\
\end{equation}
with
\begin{eqnarray}
\chi_{\alpha\beta}^{{\rm ph}}(q)&=&-\frac{T}{N}\sum_{k}G_{\alpha\beta}(k+q)G_{\beta\alpha}(k),\\
\chi_{\alpha\beta}^{{\rm pp}}(q)&=&\frac{T}{N}\sum_{k}G_{\alpha\beta}(k+q)G_{\alpha\beta}(-k).
\end{eqnarray}
Here we denote $q\equiv(\Vec{q},i\epsilon_l)$  with $\epsilon_l=2\pi lT$ 
being the Matsubara frequency for bosons, and 
$N$ the number of $\Vec{k}$-points on a mesh.

Green's function is related to $\Sigma$ via Dyson's equation,
\begin{equation}
\left[G(k)^{-1}\right]_{\alpha\beta}=\left[G^{0}(k)^{-1}\right]_{\alpha\beta}-\Sigma_{\alpha\beta}(k),\label{equend}
\end{equation}
where $G^{0}$ is the bare Green's function,
\begin{equation}
G_{\alpha\beta}^{0}(k)=\left[\frac{1}{i\omega_n+\mu-\epsilon_{\Vec{k}}^{0}}\right]_{\alpha\beta}
\end{equation}
with $\epsilon_{\Vec{k}}^{0}$ the energy of a free electron.

The gap function, $\phi$, and $T_c$ are obtained from Eliashberg's equation,
\begin{eqnarray}
\lambda\phi_{\alpha\beta}(k)&=&-\frac{T}{N}\sum_{k'}\sum_{\alpha',\beta'}\nonumber\\
& &V_{\alpha\beta}^{(2)}(k-k')G_{\alpha\alpha'}(k')G_{\beta\beta'}(-k')\phi_{\alpha'\beta'}(k')
\label{elia},
\end{eqnarray}
where the (spin-singlet) pairing interaction $V^{(2)}(q)$ is given as
\begin{eqnarray}
V_{\alpha\beta}^{(2)}(q)&=&\frac{3}{2}U^2\!\!\left[\frac{\chi^{{\rm ph}}(q)}{1-U\chi^{{\rm ph}}(q)}\right]_{\alpha\beta}\!\!\!-\frac{1}{2}U^2\!\!\left[\frac{\chi^{{\rm ph}}(q)}{1+U\chi^{{\rm ph}}(q)}\right]_{\alpha\beta}\nonumber\\
& &+U\delta_{\alpha\beta}.
\end{eqnarray}
$T_c$ is then determined as the temperature at which the largest eigenvalue, 
$\lambda$, of Eliashberg's equation becomes unity.

The spin susceptibility in RPA is given by,
\begin{equation}
\chi_{\alpha\beta}(\Vec{k},0)=\left[\frac{\chi^{{\rm ph}}(\Vec{k},0)}{1-U\chi^{{\rm ph}}(\Vec{k},0)}\right]_{\alpha\beta},
\end{equation}
may be expressed as diagonalized components,
\begin{equation}
\chi_{\pm}=\frac{\chi_{{\rm AA}}+\chi_{{\rm BB}}}{2}\pm\sqrt{\left[\frac{\chi_{{\rm AA}}-\chi_{{\rm BB}}}{2}\right]^{2}+\left|\chi_{{\rm AB}}\right|^{2}} .
\end{equation}
For the two-band FLEX calculation in 3D 
we take $N=32^3$ $\Vec{k}$-point meshes, and the
Matsubara frequencies taken as $-(2N_c-1)\pi T < \omega_n < 
(2N_c-1)\pi T$ with $N_c=4096$, which gives converged results.

\section{result}
\subsection{Stacked bond-alternating lattice}

\begin{figure}
\begin{center}
\leavevmode\epsfysize=40mm 
\epsfbox{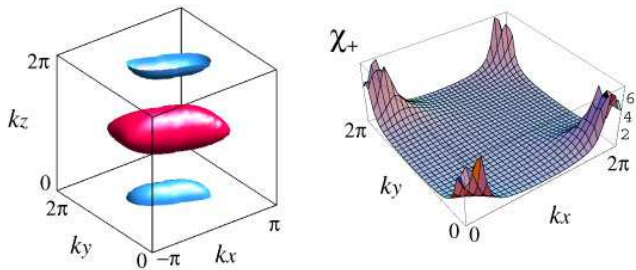}
\caption{(Color) Fermi surface (left panel; blue: bonding band, red:
antibonding band) along with $\chi_+$ at $k_z=\pi$ (right) for 
the stacked bond-alternating lattice with $U=4$, 
$n=1.02$, $t_{x_1}=0.4$, $t_{x_2}=2.0$, $t_z=0.8$, $T=0.02$.}
\label{opt-fchi}
\end{center}
\end{figure}

We have searched for the case where the inter-band nesting 
is strong with a weak intra-band nesting along the following line.  
We start from the stacked honeycomb 
lattice, where the tubular Fermi surface connected along $x-y$ directions 
has too strong an intra-band nesting as mentioned.\cite{onari}  
One way to make the Fermi surface more compact and structureless 
is to introduce alternating bonding structures in the $x-y$ plane.  
For example, if the in-plane bonds along $x$ alternate between 
weak ones ($t_{x1}$) and strong ones ($t_{x2}$), 
where the array of strong and weak ones is staggered as 
we go along $y$ direction.  We can then connect the layers 
with the third-direction hopping $t_z$, and we call this 
the stacked bond-alternating lattice (Fig. \ref{lattice}).   
We have varied 
$t_{x_1}$ and $t_z$ to examine the pairing instability 
over the region $t_{x_1},t_z<t_y(=1)<t_{x_2}$ 
with a fixed $t_{x_2}=2$ 
for the band filling $n=1.02$.  
This region has turned out to correspond to the 
case of compact Fermi pockets.  
The repulsion is taken to be $U=4$, which is 
a typical value for strongly correlated systems.  
Hereafter we take $t_y$ as the unit of energy.
The Fermi surface and the spin susceptibility for a typical 
case ($t_{x_1}=0.4$, $t_z=0.8$) is shown in Fig. \ref{opt-fchi}.
We can see that the Fermi surface comprises a 
pair of compact, ball-like pockets.  There exists, as intended, a 
dominant nesting vector in the stacking direction 
($\Vec{k}=(0,0,\pi)$) across the two pockets, 
as confirmed from a strong peak in $\chi_+(\Vec{k},0)$ around 
$\Vec{k}=(0,0,\pi)$ in Fig. \ref{opt-fchi}. 

\begin{figure}
\begin{center}
\leavevmode\epsfysize=70mm 
\epsfbox{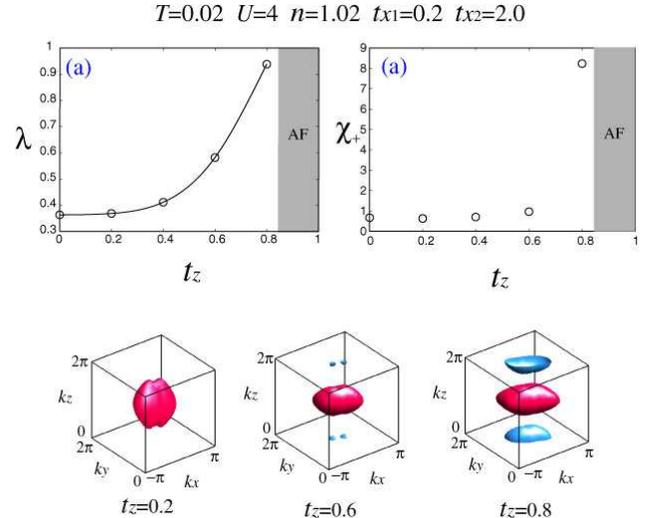}
\caption{(Color) The largest eigenvalue, $\lambda$ (top left), 
of Eliashberg's equation and the peak value of spin susceptibility 
$\chi_+$ (top right, with the gray region indicating 
the antiferromagnetic phase) 
as a function of the inter-layer 
hopping $t_z$, along with the Fermi surface 
(bottom) for selected values of $t_z$ 
for $U=4$, $n=1.02$, $t_{x_1}=0.2$, $t_{x_2}=2.0$, $T=0.02$ 
for the stacked bond-alternating lattice. 
}
\label{(a)}
\end{center}
\end{figure}

Figure \ref{(a)} shows the result for the dependence of the largest 
eigenvalue, $\lambda$, of Eliashberg's equation on the 
inter-layer hopping, $t_{z}$, for $T=0.02$ and $t_{x_1}=0.2$.  
We can see that $\lambda$ increases with $t_z$.  
This indeed occurs as the two Fermi pockets 
(displayed in red and blue) begin to be formed 
and the inter-band nesting across the pockets, that should favor the 
pairing, develops.  
For this choice of parameters, however, the superconductivity 
gives way to antiferromagnetism 
(whose boundary is identified here to the 
situation when $1-U\chi^{\rm ph}=0$) 
around the value of $t_z$ at which $\lambda$ reaches unity.  
Nevertheless we are on the right track in 
that the gap function is, as intended, nodeless (gapful) 
in each band while changes sign across the two bands. 

\begin{figure}
\begin{center}
\leavevmode\epsfysize=110mm 
\epsfbox{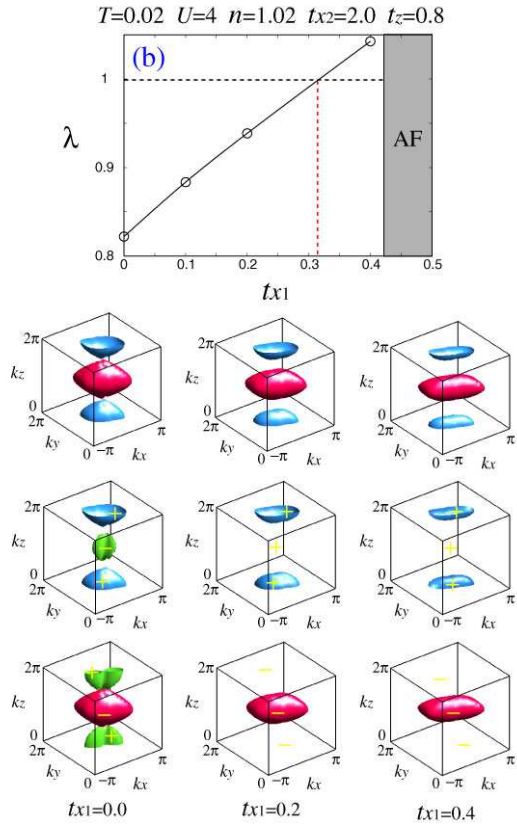}
\caption{(Color)
A plot similar to Fig. \ref{(a)} (top two rows) when we vary 
the weak hopping, $t_{x_1}$, in the stacked bond-alternating lattice 
with fixed $t_{x_2}=2.0$, $t_z=0.8$ for $U=4$, $n=1.02$,
and $T=0.02$. We depict the gap function with its sign indicated 
on the Fermi surface for the bonding (second from bottom row) and
antibonding (bottom) bands, respectively, 
where nodal planes are shown by green sheets.}
\label{(b)}
\end{center}
\end{figure}

So we should next tune the shape of the two Fermi pockets.  
This can be achieved by varying the weak hopping, $t_{x_1}$, 
in the bond-alternating system.   
As seen in Fig. \ref{(b)}, the Fermi surface for each 
band remains to be two balls throughout, but they become more compact as 
$t_{x_1}$ is increased.  
If we turn to $\lambda$ in the figure for a fixed $t_z=0.8$ with $T=0.02$, 
$\lambda$ indeed reaches unity well before the antiferromagnetism sets in.  
In the case of $t_{x1}=0$ (with a low $T_c$), 
the gap function has a nodal plane as shown by green sheet in
Fig. \ref{(b)}, which should be due to the 
main $(0,0,\pi)$ pair scattering in large
antibonding band Fermi surface favor node to change the sign across the
scattering.

So we conclude for the bond-alternating model that, 
although disconnected Fermi surfaces 
are favorable for the inter-band pairing, 
the Fermi pockets have to be compact 
so that the intra-band pair-scattering is suppressed and 
the gap function for each band becomes nodeless.

\begin{figure}
\begin{center}
\leavevmode\epsfysize=40mm 
\epsfbox{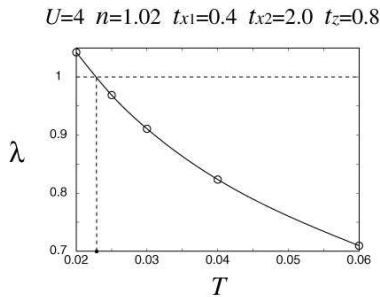}
\caption{$\lambda$ as a function of $T$ 
for the optimized $t_{x_1}=0.4$ and $t_z=0.8$ 
for the stacked bond-alternating lattice with 
$U=4$, $n=1.02$, $t_{x_2}=2.0$, and $T=0.02$.}
\label{opt-elia}
\end{center}
\end{figure}

We have then scanned the $t_{x_1}-t_z$ parameter plane to 
optimize the values of $t_{x_1}$ and $t_z$, and found 
that $\lambda$ takes its maximum at $t_{x_1}=0.4$, $t_z=0.8$. 
In Fig. \ref{opt-elia}, temperature dependence of $\lambda$ 
for the optimized case is plotted. We can see that
$T_c\sim 0.02t_y$, 
which is remarkably high (i.e., $\simeq 2\times 10^{-3} \times$ the 
bandwidth with the bandwidth $\simeq 12t_y$) for a 3D system 
(where usually $T_c\sim 10^{-4} \times$ bandwidth 
at most as estimated for the simple cubic lattice\cite{arita}).

\subsection{Stacked connected-ladder lattice}
Let us move on to the stacked connected-ladder lattice 
(right panel in Fig.\ref{lattice}).  
In this subsection, we fix $n=1.05, U=4.0, t_{x_2}=1.6, t_{y}=1({\rm unit})$.  
This value of $t_{x_2}$ turns out to give the highest $T_c$ 
for $U=4$.  In 2D Kuroki and Arita\cite{KA} have shown that 
a larger $U=8$ gives a higher $T_c$, but in the present case 
of 3D the tendency toward AF is stronger, so that 
we take a weaker $U=4$ as in the previous section. 

\begin{figure}
\begin{center}
\leavevmode\epsfysize=70mm 
\epsfbox{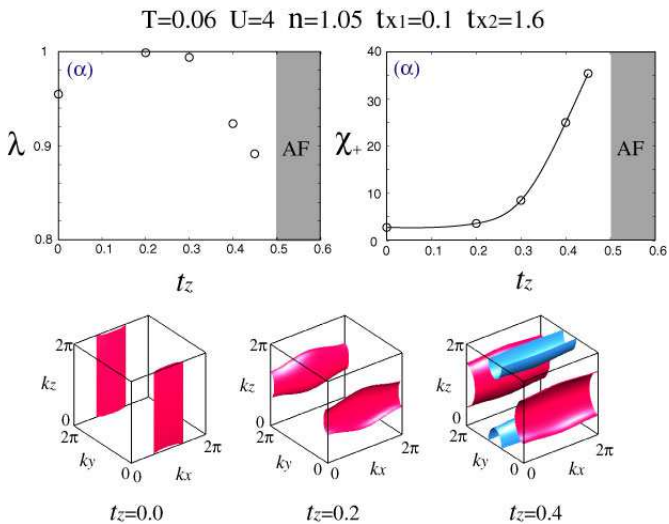}
\caption{(Color) $\lambda$ (top left) and 
$\chi_+$ (top right) as a function of the 
inter-layer hopping $t_z$ 
along with the Fermi surface for the stacked connected-ladder lattice 
with selected 
values of $t_z$ (as indicated at the bottom) for the stacked-ladder lattice 
for $U=4$, $n=1.05$, $t_{x_1}=0.1$, $t_{x_2}=1.6$, $T=0.06$. }
\label{c}
\end{center}
\end{figure}

Figure \ref{c} shows $\lambda$ and $\chi_+$ as a function of the 
inter-layer hopping $t_z$ for a fixed $t_{x1}=0.1$ with $T=0.06$. 
We can see that $\lambda$ is as large as, or even greater than, 
in 2D ($t_z=0$) up to $t_z < 0.3$. 
For smaller $t_z < 0.2$ the Fermi surface consists of 
only one band (two panels at bottom left in Fig.\ref{c}).  
In this case the inter-band nesting is weak, as seen from a small 
$\chi_+$ (top right in Fig.\ref{c}).  
When $t_z > 0.3 \gg t_{x1}$, the system becomes 
quasi-2D along the $yz$ plane, and a nesting in the direction of 
$(0,\pi,\pi)$ evolves (as evident from the shape of the Fermi cylinders 
that change their direction from $\parallel z$ to $\parallel x$).   
While $\chi_+$ monotonically increases with $t_z$ throughout, 
superconductivity is most favored ($\lambda\simeq 1$ 
in between ($t_z\simeq 0.2$).

The resultant $T_c\simeq 0.06$ in the 3D system is 
$T_c\simeq 8\times 10^{-3} \times$ 
bandwidth (with the bandwidth $\simeq 8t_y$), 
which is even higher than in the above 
bond-alternating lattice, and 
almost an order of magnitude higher than the ordinary 3D case 
of $\simeq 10^{-4} \times$ bandwidth.
This shows that the 
superconductivity in the disconnected Fermi surface 
proposed in Ref.\onlinecite{KA} for 2D is robust against 
introduction of the third-direction hopping.

\subsection{Condition favoring the pairing on disconnected Fermi surfaces}
The peak in $T_c$ at a nonzero $t_z$ (rather than at $t_z=0$) 
seen above is surprising.  
This is particularly puzzling since the spin susceptibility 
increases monotonically with $t_z$, which implies that 
a good nesting {\it per se} does not necessarily favor the pairing.  
We then want to pin-point the physical reason 
why superconductivity is degraded although $\chi_+$ increases 
in the region of $t_z>0.3$.

\begin{figure}
\begin{center}
\leavevmode\epsfysize=70mm 
\epsfbox{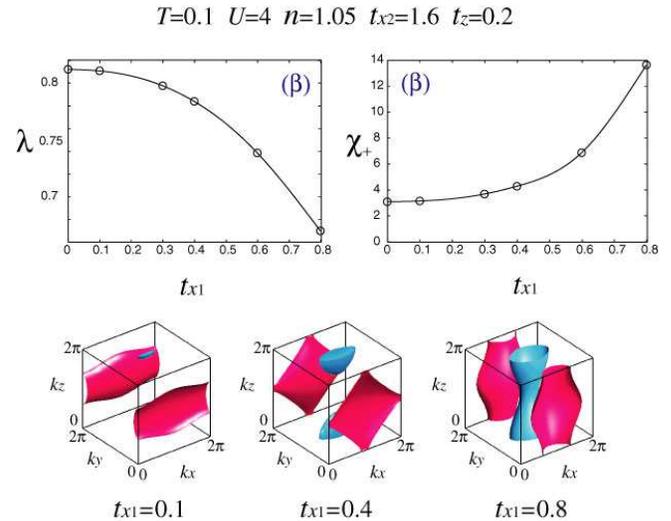}
\caption{A plot similar to Fig. \ref{c} 
for the stacked connected-ladder lattice when the 
weak hopping, $t_{x1}$, is varied as indicated at the bottom 
for a fixed $t_z=0.2$ with $U=4$, $n=1.05$, $t_{x_2}=1.6$, 
and $T=0.1$.}
\label{e}
\end{center}
\end{figure}

To elaborate the question let us 
look at another, similar phenomenon when we 
increase the in-plane, inter-ladder hopping, $t_{x_1}$, 
with a fixed $t_{z}=0.2$ in Fig. \ref{e}.  
There, we can see that, although the nesting, in the direction of 
$(0,\pi,\pi)$, evolves 
with $t_{x_1}$ as shown from both the shape of the 
Fermi surface and the value of $\chi_+$ in Fig. \ref{e},
$\lambda$ decreases as $\chi_+$ increases, which 
provides another example of the degraded 
superconductivity in the evolution of the nesting.  

\begin{figure}
\begin{center}
\leavevmode\epsfysize=40mm 
\epsfbox{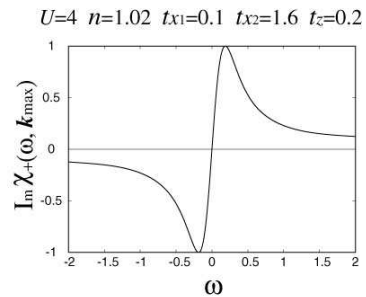}
\caption{ ${\rm Im}\chi_+(\Vec{k}_{\rm max},\omega)$ (normalized by its 
maximum value), where $\Vec{k}_{\rm max}=(0,\pi,\pi)$, 
 as a function of $\omega$ for $T=0.06, U=4, n=1.05, t_{x_1}=0.1, t_{x_2}=1.6, 
 t_z=0.2$ for the stacked connected-ladder lattice.}
\label{imchi}
\end{center}
\end{figure}

\begin{figure}
\begin{center}
\leavevmode\epsfysize=90mm 
\epsfbox{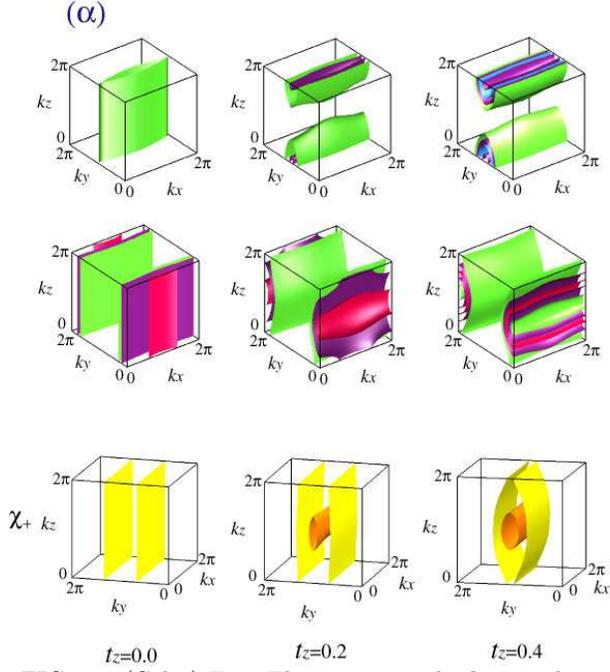}
\caption{(Color) Top: The regions in the $\Vec{k}$-space bounded by 
$\mu-\delta\mu < \epsilon^0+{\rm Re} \Sigma < \mu+\delta\mu$ 
with $\delta\mu = 0.5$(purple), $1$(green), along with the Fermi surface 
(blue: the bonding band which is only visible in the right 
row, red: the anti-bonding band) 
on which the region of $\epsilon^0+{\rm Re}\,\Sigma=\mu$ is displayed 
for the stacked connected-ladder lattice 
with selected values of the inter-layer hopping 
$t_z$  as in Fig.\ref{c} as indicated at the bottom.  
Bottom: $\Vec{k}$-space contours (orange: $\chi_+=3$, 
yellow: $\chi_+=1$) of the spin susceptibility.}
\label{d}
\end{center}
\end{figure}

\begin{figure}
\begin{center}
\leavevmode\epsfysize=90mm 
\epsfbox{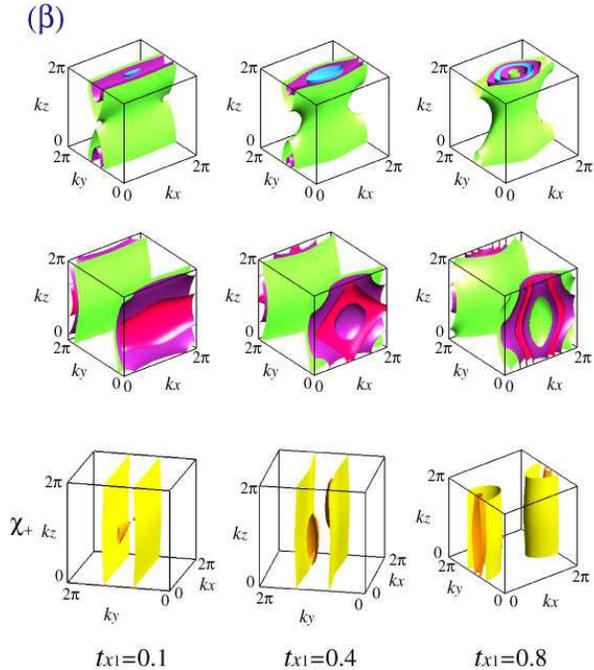}
\caption{(Color) A plot similar to Fig. \ref{d} 
for the stacked connected-ladder lattice 
when the weak hopping, $t_{x_1}$, is varied as in Fig.\ref{e} 
as indicated at the bottom.}
\label{f}
\end{center}
\end{figure}

\begin{figure}
\begin{center}
\leavevmode\epsfysize=35mm 
\epsfbox{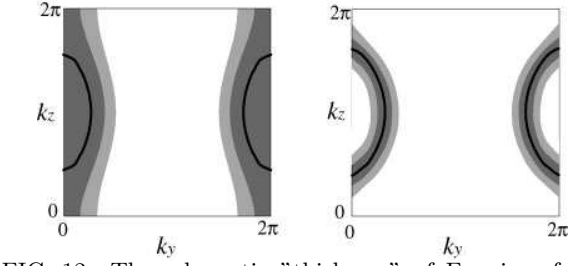}
\caption{The schematic ''thickness'' of Fermi surface at $k_x=\pi$. The light gray and dark gray regions
 represent $0.5<|\epsilon^0-\mu+{\rm Re} \Sigma|<1$ and
 $0<|\epsilon^0-\mu+{\rm Re} \Sigma|<0.5$, respectively, and Fermi
 surfaces are shown by black lines. The left panel has larger gray
 (active) region than the right.}
\label{cont}
\end{center}
\end{figure}

We can conceive several reasons as follows.  
As we have mentioned, the increase of $t_z$ or $t_{x1}$ 
raises the dimensionality of the anisotropic system 
to 3D.  This should degrade superconductivity 
in two ways. First one is just the mechanism revealed 
by Arita {\it et al.} in ref.\cite{arita}, where 
higher the dimension smaller the 
phase volume fraction for the pairing interaction.  
Secondly, when $E_F$ is close to band edges, 
the density of states there tends to be smaller for higher dimensions. 
To see that this is indeed the case, 
we have examined in Figs.\ref{d},\ref{f} the regions in $\Vec{k}$-space in which 
\[
\mu-\delta\mu < \epsilon^0+{\rm Re}\Sigma < \mu+\delta\mu, 
\]
which we will refer to as the ``thickness'' of the Fermi surface.  
Bulkier the thickness larger the density of states, 
but we specifically take the energy window $\delta\mu$
as the energy interval over which 
${\rm Im}\chi_+({\Vec{k}_{\rm max}},\omega)$
(with $\Vec{k}_{\rm max}$ being the momentum that gives 
the peak in $\chi_+$), displayed in Fig.\ref{imchi}, is significant.
In other words, $\delta\mu$ is the energy scale of the spin fluctuations 
that mediate the pairing.  
As seen in Figs. \ref{d},\ref{f}, respectively, 
increased $t_z$ or $t_{x1}$ makes the 
Fermi surface ``thinner" although the Fermi surface itself is large.  
The situation is shown schematically in Fig.\ref{cont}

Another factor governing superconductivity 
can also be identified from Figs.\ref{d},\ref{f}.  
Namely, since we are here dealing with small (or vanishing)
Fermi surfaces, the spin susceptibility is governed not by the shape 
of the Fermi surface itself, but by the thickened Fermi surface. 
The thin Fermi surfaces  
for smaller $t_z$ or $t_{x1}$ then make the nesting sharper and 
the peak in $\chi_{+}$ becomes narrower. 
This also acts to degrade superconductivity, since 
the spin susceptibility has to be appropriately ``blurred''
in order to make the spin fluctuation usable over 
a large phase space for pair scattering.

\section{Conclusion}
We have studied how the pairing with higher $T_c$ in disconnected 
Fermi surfaces can be realized in three-dimensional systems.  
Our first proposal of the stacked bond-alternating lattice has a 
pair of ball-like Fermi pockets with a dominant inter-band nesting vector and
antiferromagnetic spin fluctuations, which is a genuinely 
3D model in that the interlayer 
hopping $t_z$ promotes the inter-band pairing. 
$T_c$  estimated from Eliashberg's equation shows that 
$T_c \sim 10^{-3}$ times the bandwidth ($\sim O(0.01t)$), 
which is as high as that for the single-band 
Hubbard model on the square lattice, and remarkably high for 3D system.

Our second model of the stacked layers each having 
a disconnected Fermi-surface in 2D originally proposed 
by Kuroki and Arita\cite{KA} is shown to have a high $T_c$ that is 
robust against the introduction of the inter-layer hopping, 
although the hopping degrades the 2D nesting. The estimated 
$T_c \sim$ several times $10^{-3} \times$ bandwidth ($=$ 
several times $O(0.01t)$) is even higher than the first model.  

From these we have then identified the right conditions for 
higher $T_c$ in the disconnected Fermi surface: 
the system should be quasi-low dimensional, 
and the peak in the spin susceptibility should be not too sharp but 
``blurred''.  So the message here is 
that 3D materials with considerably high $T_c$ can be 
expected if we consider appropriate lattice structures.  

\section{Acknowledgements}
Numerical calculations were performed at the Computer Center and the
ISSP Supercomputer Center of University of Tokyo. This study was in part
supported by a Grant-in-aid for scientific research from the Ministry of
Education of Japan.

\end{multicols}
\end{document}